\documentclass[a4paper,12pt]{article}
\usepackage{epsfig}
\usepackage{feynmf}
\usepackage{a4p}
\usepackage{cite}
\def\ks{\mbox{K}^0_{\rm S}}
\def\Ks{\mbox{K}^0_{\rm S}}
\def\gg{\gamma\gamma}
\def\ee{\mbox{e}^+\mbox{e}^-}

\def\sqee{\sqrt{s}_{\rm ee}}
\def\WECAL{W_{\rm ECAL}}

\def\pT{p_{\rm T}}
\def\pt{p_{\rm T}}

\def\fg{f_{\gamma/{\rm e}}}

\def\qmax{Q^2_{\rm max}}
\def\qmin{Q^2_{\rm min}}
\def\qqbar{\mbox{q}\overline{\mbox{q}}}
\def\dspt{{\rm d}\sigma/{\rm d}p_{\rm T}} 
\def\dseta{{\rm d}\sigma/{\rm d}|\eta|} 
 
\def\pz{\phantom{0}}
\def\pzz{\phantom{00}}
\newcommand{\sleq} {\raisebox{-.6ex}{${\textstyle\stackrel{<}{\sim}}$}}

\begin{document}
\renewcommand{\thefootnote}{\arabic{footnote}}
\renewcommand{\arraystretch}{1.0}
\renewcommand{\Huge}{\huge}

\begin{titlepage}
\begin{center}{\large   EUROPEAN LABORATORY FOR PARTICLE PHYSICS
}\end{center}\bigskip
\begin{flushright}
       CERN-EP/98-091   \\ 4 June 1998
\end{flushright}
\bigskip\bigskip\bigskip\bigskip\bigskip
\begin{center}{\LARGE\bf   
\boldmath Inclusive Production of Charged
Hadrons and $\ks$ Mesons 
in Photon-Photon Collisions \unboldmath
}\end{center}\bigskip\bigskip
\begin{center}{\LARGE The OPAL Collaboration
}\end{center}\bigskip\bigskip
\begin{center}
%
\end{center}
\bigskip\begin{center}{\large  Abstract}\end{center}
The production of charged hadrons and $\ks$ mesons in the collisions
of quasi-real photons has been measured 
using the OPAL detector at LEP. The data were taken at
$\ee$ centre-of-mass energies of $161$ and $172$~GeV.
The differential cross-sections as a function of the 
transverse momentum and the pseudorapidity of the charged 
hadrons and $\ks$ mesons
have been compared to the leading order 
Monte Carlo simulations of PHOJET and PYTHIA and 
to perturbative next-to-leading order (NLO) QCD calculations. 
The distributions have been measured in the range $10<W<125$~GeV
of the hadronic invariant mass $W$.
By comparing the transverse momentum distribution of charged hadrons measured
in $\gg$ interactions with $\gamma$-proton and meson-proton data
we find evidence for hard photon interactions in addition
to the purely hadronic photon interactions.
\bigskip\bigskip\bigskip\bigskip
\bigskip\bigskip
\begin{center}{\large
(submitted to European Physics Journal C)
}\end{center}
\end{titlepage}
\begin{center}{\Large        The OPAL Collaboration
}\end{center}\bigskip
\begin{center}{
K.\thinspace Ackerstaff$^{  8}$,
G.\thinspace Alexander$^{ 23}$,
J.\thinspace Allison$^{ 16}$,
N.\thinspace Altekamp$^{  5}$,
K.J.\thinspace Anderson$^{  9}$,
S.\thinspace Anderson$^{ 12}$,
S.\thinspace Arcelli$^{  2}$,
S.\thinspace Asai$^{ 24}$,
S.F.\thinspace Ashby$^{  1}$,
D.\thinspace Axen$^{ 29}$
% ...
}\end{center}\bigskip
\bigskip
$^{  1}$School of Physics and Astronomy, University of Birmingham,
Birmingham B15 2TT, UK
\newline
$^{  2}$Dipartimento di Fisica dell' Universit\`a di Bologna and INFN,
I-40126 Bologna, Italy
\newline
$^{  3}$Physikalisches Institut, Universit\"at Bonn,
D-53115 Bonn, Germany
\newline
$^{  4}$Department of Physics, University of California,
Riverside CA 92521, USA
\newline
$^{  5}$Cavendish Laboratory, Cambridge CB3 0HE, UK
\newline
$^{  6}$Ottawa-Carleton Institute for Physics,
Department of Physics, Carleton University,
Ottawa, Ontario K1S 5B6, Canada
\newline
$^{  7}$Centre for Research in Particle Physics,
Carleton University, Ottawa, Ontario K1S 5B6, Canada
\newline
$^{  8}$CERN, European Organisation for Particle Physics,
CH-1211 Geneva 23, Switzerland
\newline
$^{  9}$Enrico Fermi Institute and Department of Physics,
University of Chicago, Chicago IL 60637, USA
\newline
$^{ 10}$Fakult\"at f\"ur Physik, Albert-Ludwigs-Universit\"at,
D-79104 Freiburg, Germany
\newline
$^{ 11}$Physikalisches Institut, Universit\"at
Heidelberg, D-69120 Heidelberg, Germany
\newline
$^{ 12}$Indiana University, Department of Physics,
Swain Hall West 117, Bloomington IN 47405, USA
\newline
$^{ 13}$Queen Mary and Westfield College, University of London,
London E1 4NS, UK
\newline
$^{ 14}$Technische Hochschule Aachen, III Physikalisches Institut,
Sommerfeldstrasse 26-28, D-52056 Aachen, Germany
\newline
$^{ 15}$University College London, London WC1E 6BT, UK
\newline
$^{ 16}$Department of Physics, Schuster Laboratory, The University,
Manchester M13 9PL, UK
\newline
$^{ 17}$Department of Physics, University of Maryland,
College Park, MD 20742, USA
\newline
$^{ 18}$Laboratoire de Physique Nucl\'eaire, Universit\'e de Montr\'eal,
Montr\'eal, Quebec H3C 3J7, Canada
\newline
$^{ 19}$University of Oregon, Department of Physics, Eugene
OR 97403, USA
\newline
$^{ 20}$Rutherford Appleton Laboratory, Chilton,
Didcot, Oxfordshire OX11 0QX, UK
\newline
$^{ 22}$Department of Physics, Technion-Israel Institute of
Technology, Haifa 32000, Israel
\newline
$^{ 23}$Department of Physics and Astronomy, Tel Aviv University,
Tel Aviv 69978, Israel
\newline
$^{ 24}$International Centre for Elementary Particle Physics and
Department of Physics, University of Tokyo, Tokyo 113, and
Kobe University, Kobe 657, Japan
\newline
$^{ 25}$Institute of Physical and Environmental Sciences,
Brunel University, Uxbridge, Middlesex UB8 3PH, UK
\newline
$^{ 26}$Particle Physics Department, Weizmann Institute of Science,
Rehovot 76100, Israel
\newline
$^{ 27}$Universit\"at Hamburg/DESY, II Institut f\"ur Experimental
Physik, Notkestrasse 85, D-22607 Hamburg, Germany
\newline
$^{ 28}$University of Victoria, Department of Physics, P O Box 3055,
Victoria BC V8W 3P6, Canada
\newline
$^{ 29}$University of British Columbia, Department of Physics,
Vancouver BC V6T 1Z1, Canada
\newline
$^{ 30}$University of Alberta,  Department of Physics,
Edmonton AB T6G 2J1, Canada
\newline
$^{ 31}$Duke University, Dept of Physics,
Durham, NC 27708-0305, USA
\newline
$^{ 32}$Research Institute for Particle and Nuclear Physics,
H-1525 Budapest, P O  Box 49, Hungary
\newline
$^{ 33}$Institute of Nuclear Research,
H-4001 Debrecen, P O  Box 51, Hungary
\newline
$^{ 34}$Ludwigs-Maximilians-Universit\"at M\"unchen,
Sektion Physik, Am Coulombwall 1, D-85748 Garching, Germany
\newline
\bigskip\newline
$^{  a}$ and at TRIUMF, Vancouver, Canada V6T 2A3
\newline
$^{  b}$ and Royal Society University Research Fellow
\newline
$^{  c}$ and Institute of Nuclear Research, Debrecen, Hungary
\newline
$^{  d}$ and Department of Experimental Physics, Lajos Kossuth
University, Debrecen, Hungary
\newline
$^{  e}$ on leave of absence from the University of Freiburg
\newline
\section{Introduction}
Inclusive hadron production in collisions of quasi-real photons
can be used to study the structure of photon interactions
complementing similar studies of jet production in $\gg$
collisions~\cite{bib-opalgg}.
The photons are radiated by the beam electrons\footnote{Positrons
are also referred to as electrons} carrying only
small negative
squared four-momenta $Q^2$. They can therefore be considered to be
quasi-real ($Q^2 \approx 0$) if the
electrons are scattered at very small angles where they
are not detected. For the ``anti-tagged'' event sample, events
are rejected if one or both scattered electrons have been detected.

The interactions of the photons can be modelled by assuming that
each photon can either interact directly or appear resolved
through its fluctuations into hadronic states. 
In leading order Quantum Chromodynamics (QCD) this 
model leads to three different event classes for the $\gamma\gamma$ 
interactions: direct, single-resolved and 
double-resolved. In resolved events partons 
(quarks or gluons) from the hadronic fluctuation of the photon
take part in the hard interaction. 
The probability to find a parton in the photon 
carrying a certain momentum fraction of the photon is parametrised
by parton density functions.

We measure differential production cross-sections
as a function of the transverse momentum and the pseudorapidity
of charged hadrons and neutral $\ks$ mesons.
Since the distributions are fully corrected for
losses due to event and track selection cuts, the acceptance and
the resolution of the detector, they are directly comparable 
to leading order Monte Carlo models
and to next-to-leading order (NLO) perturbative QCD calculations by
Binnewies, Kniehl and Kramer~\cite{bib-binnewies}. 
Until now, transverse momentum distributions of charged hadrons have only been
measured for single-tagged events by TASSO~\cite{bib-tasso} and
MARK~II~\cite{bib-mark2} at an average $\langle Q^2 \rangle$
of 0.35~GeV$^2$ and 0.5~GeV$^2$, respectively.
We present the first measurement
in anti-tagged collisions of quasi-real photons. Furthermore, 
the transverse momentum distributions in $\gg$ interactions are expected to
have a harder component than in photon-proton or meson-proton
interactions due to the direct photon interactions. This will be
demonstrated by comparing our data to the photo- and hadroproduction 
data measured by WA69~\cite{bib-wa69}.

At large transverse momenta (after crossing the charm threshold) 
the production of $\ks$ mesons in photon-photon collisions
is sensitive to the direct production of primary charm quarks
in addition to the production of primary strange quarks, since
the photon couples to the quark charge.
$\ks$ production in anti-tagged $\gg$ collisions has 
previously been measured by TOPAZ~\cite{bib-topaz} and 
in single-tagged events by MARK~II~\cite{bib-mark2}.

In this paper, charged hadron and $\ks$ production are studied
using the full data sample taken in 1996 at $\ee$ centre-of-mass
energies of 161 and 172~GeV corresponding to an
integrated luminosity of about 20~pb$^{-1}$.

\section{The OPAL detector}
\label{sec-dec}
A detailed description of the OPAL detector
can be found in Ref.~\cite{opaltechnicalpaper}, and
therefore only a brief account of the main features relevant
to the present analysis will be given here.
 
The central tracking system is located inside 
a solenoidal magnet which
provides a uniform axial magnetic field of 0.435~T along the beam
axis\footnote{In the OPAL coordinate system 
  the $z$ axis points in the direction of the e$^-$ beam. The
  polar angle $\theta$, the azimuthal angle $\phi$
  and the radius $r$ denote the usual spherical coordinates.}.
The detection efficiency for charged particles is close to 100~$\%$
within the polar angle range $|\cos\theta|<0.92$.
The magnet is surrounded in the barrel region ($|\cos\theta|<0.82$)
by a lead glass electromagnetic
calorimeter (ECAL) and a hadronic sampling calorimeter (HCAL).  
Outside the HCAL, the detector is surrounded by muon
chambers. There are similar layers of detectors in the 
endcaps ($0.81<|\cos\theta|<0.98$). 
The small angle region from 47 to 140 mrad
around the beam pipe on both sides
of the interaction point is covered by the forward calorimeters (FD)
and the region from 25 to 59 mrad by the silicon tungsten luminometers (SW).
From 1996 onwards, including the data presented in this paper,
the lower boundary of the acceptance has been  increased to 33 mrad
following the installation of a low angle shield to protect the
central detector against possible synchrotron radiation.
 
Starting with the innermost components, the
tracking system consists of a high precision silicon
microvertex detector, a vertex
drift chamber, a large volume jet chamber with 159 layers of axial
anode wires and a set of $z$ chambers measuring the track coordinates
along the beam direction. 
The transverse momenta $\pt$ of tracks are measured with a precision 
parametrised by
$\sigma_{\pt}/\pt=\sqrt{0.02^2+(0.0015\cdot \pt)^2}$ ($\pt$ in GeV/$c$)
in the central region. In this paper ``transverse''
is always defined with respect to the $z$ axis.
The jet chamber also provides 
measurements of the energy loss, ${ \rm d} E/ {\rm d}x$, 
which are used for particle identification~\cite{opaltechnicalpaper}.

The barrel and endcap sections of the ECAL  are
both constructed from lead glass blocks with a depth of
$24.6$ radiation lengths in the barrel region and more than 
$22$ radiation lengths in the endcaps. 
The FD consist of cylindrical lead-scintillator calorimeters with a depth of   
24 radiation lengths divided azimuthally into 16 segments.  
The electromagnetic energy resolution is about
$18\%/\sqrt{E}$, where $E$ is in GeV.                                  
The SW detectors~\cite{bib-siw} consist
of 19 layers of silicon detectors and 18
layers of tungsten, corresponding to a total of 22 radiation
lengths. Each silicon layer consists of 16 wedge
shaped silicon detectors. The electromagnetic energy resolution is about
$25\%/\sqrt{E}$ ($E$ in GeV).  

\section{Kinematics and Monte Carlo simulation}
The properties of the two interacting photons ($i=1,2$)
are described by their negative four-momentum transfers $Q_{i}^2$.
Each $Q_i^2$ is related to the electron
scattering angle $\theta'_i$ relative to the beam direction by
\begin{equation}
Q_i^2=-(p_i-p'_i)^2\approx 2E_i E'_i(1-\cos\theta'_i),
\label{eq-q2}
\end{equation}
where $p_i$ and $p'_i$ are the four-momenta of the beam
electrons and the scattered electrons, respectively,
and $E_i$ and $E'_i$ are their energies.
Events with detected scattered electrons (single-tagged or
double-tagged events) are excluded from
the analysis. This anti-tagging condition 
is met when the scattering angle $\theta'$ of
the electron is less than 33~mrad between the beam axis and
the inner edge of the SW detector. 
It defines an effective upper
limit, $Q^2_{\rm max}$, on the values of $Q_{i}^2$ for both photons.
The hadronic final state is described by its invariant mass $W$.
The spectrum of photons with an energy fraction $y$ of 
the electron beam may be obtained
by the Equivalent Photon Approximation (EPA)~\cite{bib-wwa}:
\begin{equation}
\fg(y)=
\frac{\alpha}{2\pi}\left(\frac{1+(1-y)^2}{y}
\log\frac{\qmax}{\qmin}
-2m^2_{\rm e} y\left( \frac{1}{\qmin}-\frac{1}{\qmax} \right)\right),
\end{equation}
with $\alpha$ being the electromagnetic coupling constant.
The minimum kinematically allowed negative squared four-momentum 
transfer $\qmin$ is determined by the electron mass $m_{\rm e}$: 
\begin{equation}
\qmin=\frac{m_{\rm e}^2y^2}{1-y}.
\end{equation}
The Monte Carlo generators
PYTHIA~\cite{bib-pythia} and PHOJET~\cite{bib-phojet} have been 
used to simulate quasi-real photon-photon interactions.
More details about the event generation can be found in Ref.~\cite{bib-opalgg}.
All possible hard interactions relevant to photon-photon
interactions are included. The fragmentation is handled by 
JETSET~\cite{bib-pythia}.
PYTHIA uses the SaS-1D parametrisation~\cite{bib-sas1d} for the 
parton densities of the photon and PHOJET uses the GRV 
parametrisation~\cite{bib-grv}. An approximation is used for
the processes with primary charm quarks, i.e. where
the charm quark is produced in the hard interaction.
These processes are simulated
using the matrix elements for light quarks.
Subsequently the charm quarks are put on the mass-shell.


\section{Event selection and background}
\label{sec-evsel}
The production of charged hadrons and $\ks$ mesons was studied
using the data taken at $\ee$ centre-of-mass energies, $\sqee$, 
of 161 and 172~GeV with an integrated luminosity of about 9.9~pb$^{-1}$
and 10.0~pb$^{-1}$, respectively.
Photon-photon events are selected with the following set of cuts:
\begin{itemize}
\item
The sum of all energy deposits
in the ECAL and the HCAL has to be less than 45 GeV.
\item
The visible invariant hadronic mass, $\WECAL$, calculated
from the position and the energy of the clusters measured
in the ECAL, has to be greater than 3 GeV.
\item 
The missing transverse energy of the event
measured in the ECAL and the forward
calorimeters has to be less than 5 GeV.
\item 
At least 3 tracks must have been found in the tracking chambers.
A track is required to have a minimum transverse momentum
of 120 MeV/$c$,
at least 20 hits in the central jet chamber,
and the innermost hit of the track must be within a radius of 60 cm
with respect to the $z$ axis.
The distance of the point of closest approach to the origin
in the r$\phi$ plane must be
less than 30 cm in the $z$ direction and less than
2 cm in the $r\phi$~plane. 
Tracks with a momentum error larger than the momentum itself
are rejected if they have fewer than 80 hits. 
The number of measured
hits in the jet chamber must be more than half of the number of possible hits.
The number of possible hits
is calculated from the polar angle $\cos\theta$ of the track, assuming 
that the track has no curvature. 
\item
To remove events with scattered electrons in the FD or SW,
the total energy measured in the FD has to be less than
50 GeV and the total energy measured in the SW
has to be less than 35 GeV.
These cuts also reduce contamination from multihadronic events with
their thrust axis close to the beam direction.
\item
To reduce the background due to beam-gas and beam-wall interactions, 
$|\langle z_{0}\rangle|$ must be smaller than 10 cm where
$\langle z_{0}\rangle$ is the error-weighted
average of the track's $z$ coordinates at the point of closest approach to
the origin in the $r\phi$~plane.
Beam-wall events with a vertex in the beam-pipe are rejected
by requiring the radial position of the primary vertex in the $r\phi$~plane
to be less than 3~cm.
\end{itemize}
After all cuts 56732 events remain.

All relevant background processes 
apart from beam-gas and beam-wall events
were studied using Monte Carlo
generators. Multihadronic events ($\ee\rightarrow \qqbar(\gamma)$) were
simulated with PYTHIA 5.722~\cite{bib-pythia}.
KORALZ 4.02~\cite{bib-koralz} was used
to generate the process $\ee\rightarrow\tau^+\tau^-(\gamma)$ and
BHWIDE~\cite{bib-bhwide} to generate the Bhabha process 
$\ee\rightarrow\ee(\gamma)$. 
Processes with four fermions in the final state, including 
W pair production, were simulated with grc4f~\cite{bib-grc4f},
EXCALIBUR~\cite{bib-excal},
VERMASEREN~\cite{bib-vermaseren} and FERMISV~\cite{bib-fermisv}.
All signal and background Monte Carlo samples were generated
with a full simulation of the OPAL detector~\cite{bib-gopal}.
They were analysed using the same reconstruction algorithms as for the data.
The main background processes are multi-hadronic $\ee$ annihilation events and 
$\ee\rightarrow\ee\tau^+\tau^-$ events.
Other background processes are found to be negligible.
The multihadronic background is mainly reduced by the cut on the sum of the
energy measured by the HCAL and the ECAL and by the cut on the 
missing transverse energy.
The background from all these processes after the selection cuts 
amounts to less than $1\%$.

The cut on the energy in SW and FD
rejects photon-photon events with electrons scattered
at angles $\theta'$ larger than 33 mrad and with an energy greater than 35~GeV
in the SW or greater than 50~GeV in the FD.
From the Monte Carlo, the rate of events
with $\theta'>33$~mrad and energies less than 50 GeV
is estimated to be negligible. The effective anti-tagging
condition is therefore $\theta'<33$~mrad.
%

\section{Analysis}
\subsection{Correction procedure}
The measured transverse momentum and pseudorapidity
distributions of the charged hadrons and the
$\ks$ mesons have to be corrected for
losses due to the event and track selection cuts, for the acceptance and
for the resolution of the detector. This is done with 
Monte Carlo events which were generated with PYTHIA 5.722 and
PHOJET 1.05c. The data are corrected
by multiplying the experimental distribution, e.g.~of
the transverse momentum $\pT$, with correction
factors which are calculated as the bin-by-bin
ratio of the generated and the reconstructed Monte Carlo
distributions: 
\begin{equation}
\left(\frac{{\rm d}\sigma}{{\rm d} \pT}
\right)_{\rm corrected}=
\frac{\left(\frac{{\rm d}\sigma}{{\rm d} \pT}
\right)^{\rm MC}_{\rm generated~~~~}}{
\left(\frac{{\rm d}\sigma}{{\rm d} \pT}
\right)^{\rm MC}_{\rm reconstructed}}
\left(\frac{{\rm d}\sigma}{{\rm d} \pT}
\right)_{\rm measured}.
\label{eq1}
\end{equation}
As a correction factor the mean value from PYTHIA and PHOJET
is used.
The distributions of the pseudorapidity $\eta=-\ln\tan(\theta/2)$ are 
corrected in the same way. 
This method only yields reliable results if the migration 
between bins due to the finite resolution is
small. The bins of the $\pt$ and $|\eta|$ distributions have therefore 
been chosen to be significantly larger than the resolution expected from
the Monte Carlo simulation. 
The average transverse
momentum, $\langle \pt \rangle$, and the average pseudorapidity, 
$\langle |\eta| \rangle$, in each bin 
is calculated directly from the data, since detector corrections
are small compared to the statistical errors. 

\begin{figure}[htbp]
   \begin{center}
      \mbox{
          \epsfxsize=16.5cm
          \epsffile{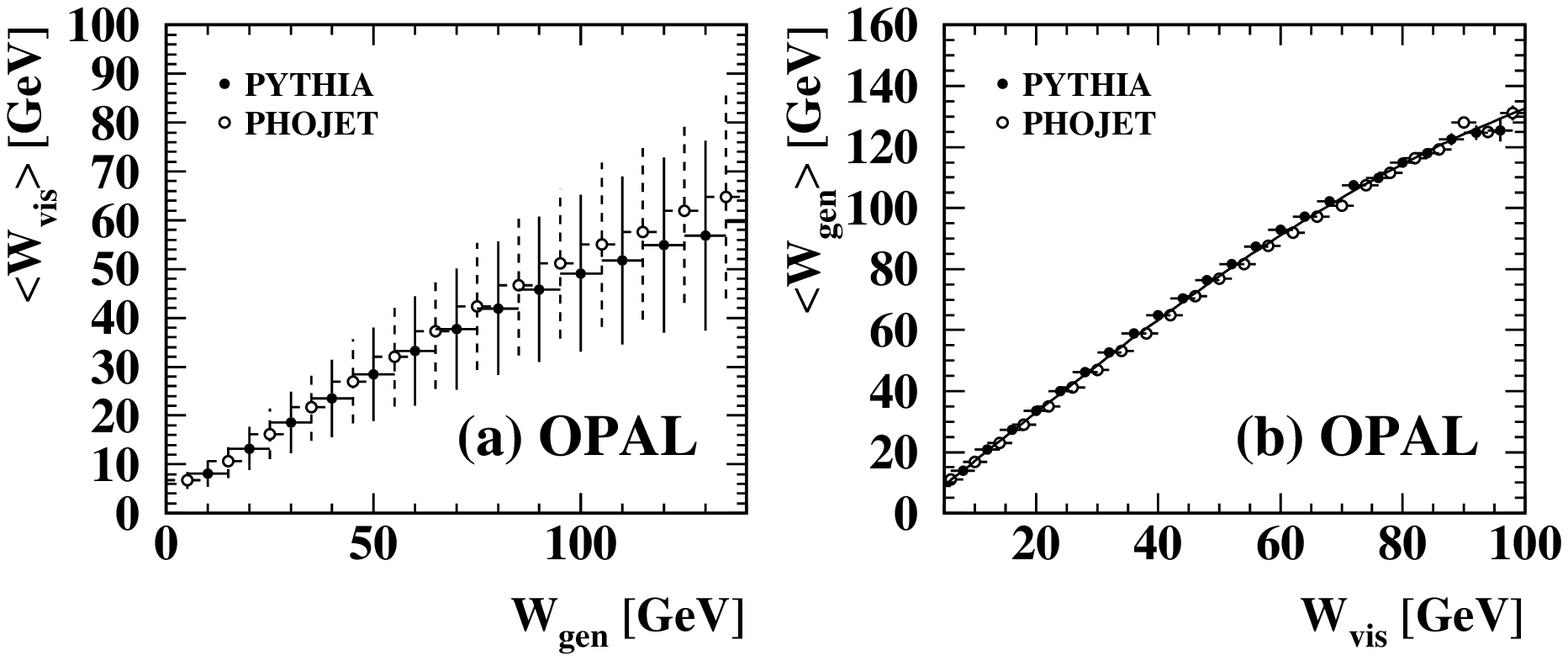}
           }
   \end{center}
\caption{The relation between the generated hadronic invariant mass 
$W_{\rm gen}$ and the visible hadronic invariant mass $W_{\rm vis}$ for
PHOJET and PYTHIA Monte Carlo events. 
The vertical bars show the standard deviation in each bin in (a) and
the error on the mean in (b). The polynomial fit shown in (b)
determines the correction function for $W_{\rm vis}$.
}
\label{fig-wcorr}
%
\bigskip
   \begin{center}
      \mbox{
          \epsfxsize=8.0cm
          \epsffile{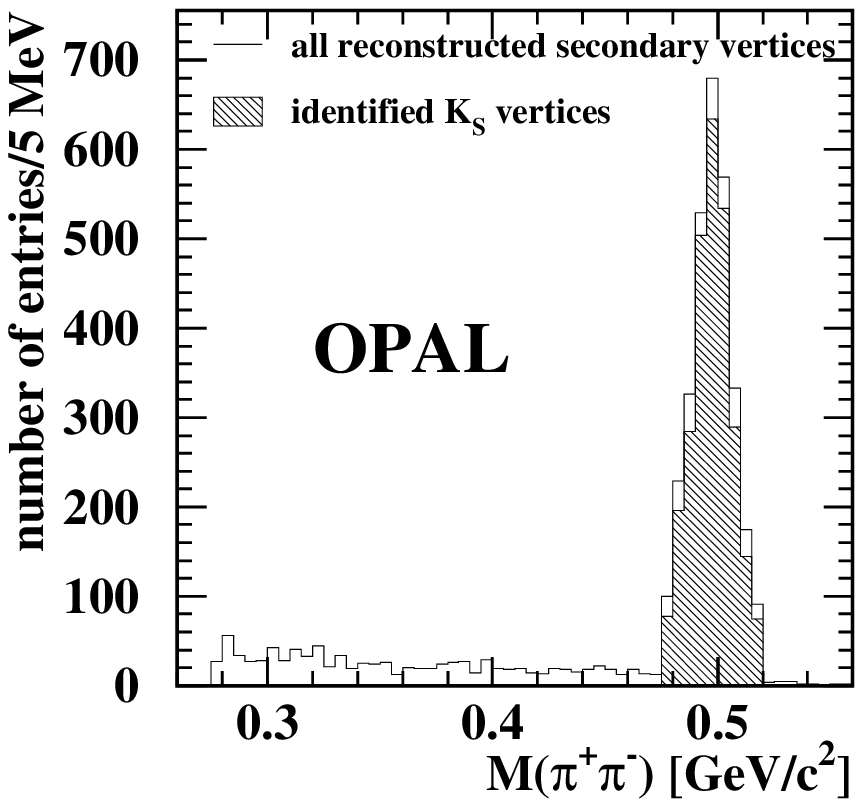}
           }
   \end{center}
\caption{
Distribution of the $\pi^+\pi^-$ invariant mass $M$ 
for all identified secondary vertices in the selected
events before (open histogram) and after (hatched histogram) 
applying the cut $|M(\pi^+\pi^-)-0.4977\mbox{~GeV}/c^2|< 0.02$~GeV$/c^2$ 
and the ${\rm d} E/{\rm d}x$ cuts
used to identify $\ks\rightarrow\pi^+\pi^-$ vertices.
}
\label{fig-mpp}
\end{figure}

The visible invariant mass, $W_{\rm vis}$, is determined from all tracks
and calorimeter clusters, including the FD and
the SW detectors. An algorithm is applied 
to avoid double-counting of particle momenta
in the central tracking system and the calorimeters~\cite{bib-opalgg}. 
All distributions are shown for $10<W<125$~GeV where
$W$ is the hadronic invariant mass corrected for
detector effects. 
To minimize migration effects when using Eq.~\ref{eq1}
for the detector correction, the bins in
$W$ must be larger than the experimental resolution
and the average reconstructed hadronic invariant mass, 
$\langle W_{\rm rec}\rangle$, should be approximately equal to the 
average generated hadronic invariant mass, $\langle W_{\rm gen}\rangle$. 
The average $\langle W_{\rm vis}\rangle $ and the resolution
on $W_{\rm vis}$ as a function of the generated hadronic invariant mass 
$W_{\rm gen}$ are therefore shown in Fig.~\ref{fig-wcorr}a, where
the vertical bars show the standard deviation (resolution) in each bin.
The average $\langle W_{\rm gen}\rangle $ as a function of  
$W_{\rm vis}$ is plotted in 
Fig.~\ref{fig-wcorr}b, where the vertical bars give the
error on the mean. This plot is used to determine
a correction function so that 
$\langle W_{\rm gen}\rangle/W_{\rm rec} \approx 1$.
The value of $W_{\rm vis}$ measured in the detector is on average
significantly smaller than $W_{\rm gen}$. 
The relation between $W_{\rm gen}$ and
$W_{\rm vis}$ shown in Fig.~\ref{fig-wcorr}b is almost
independent of the beam energy and the Monte Carlo generator used.
A single polynomial is 
therefore used to calculate $W_{\rm rec}$ from $W_{\rm vis}$.
The polynomial is obtained from the fit shown
in Fig.~\ref{fig-wcorr}b. It is applied to the data and the Monte Carlo. 

The efficiency to reconstruct photon-photon events in the detector,
estimated by the Monte Carlo,
is greater than $20\%$ for $W_{\rm gen}>10$~GeV and 
greater than $60\%$ for $W_{\rm gen}>50$~GeV.
The trigger efficiency is defined as the ratio of the number
of selected and triggered events to the number of selected events.
It was studied using data samples which
were obtained using nearly independent sets of triggers.
On average the trigger efficiency for the lowest $W$ range, $10<W<30$~GeV,
is greater than $97\%$ and it approaches $100\%$ for larger values of $W$.
Only lower limits on the trigger efficiency can be determined with
this method and therefore no correction factor is applied.

\subsection{Charged hadron production}

For the charged hadron analysis 
only particles with a proper lifetime $\tau>0.3$~ns
are used to define the primary charged hadronic multiplicity
in the Monte Carlo. 
The primary charged hadrons originate either directly from the 
primary interaction or from the decay of particles with a lifetime 
$\tau<0.3$~ns including $\Lambda$ and $\ks$ decay products.
The track selection criteria are defined as in Section~\ref{sec-evsel}.
In order to avoid regions where the detector has little or no
acceptance, all measurements of charged hadrons 
were restricted to the range $|\eta|<1.5~(|\cos\theta|\sleq~0.9)$.
In this range, the resolution on $\pt$ is given by
$\sigma_{\pt}/\pt \approx 0.02$ (see Section~\ref{sec-dec}) and
the resolution on $\eta$ by $\sigma_{\eta}\approx 0.02$.
For the $\pt$ distribution in the range $10<W<125$~GeV 
the correction factors as defined in Eq.~\ref{eq1} decrease from about 
1.7 for $\pt>120$~MeV/$c$ to about $1.1-1.4$ for $\pt>2$~GeV/$c$.
The correction factor of about 1.6 for the $\eta$ distribution is nearly 
constant for $|\eta|<1.5$.
The PHOJET and PYTHIA correction factors differ by about $3-10~\%$.

\subsection{K\boldmath $^0_{\rm S}$  production}
The $\ks$ mesons are reconstructed using
the decay channel $\ks\rightarrow \pi^+\pi^-$ which has a branching ratio
of about 69$\%$~\cite{bib-pdg}. The reconstruction procedure is similar to 
the procedure described in Ref.~\cite{bib-z01995}. It has been
optimised to increase the efficiency for finding $\ks$ mesons in photon-photon
events.
Tracks of opposite charge are paired together. 
In addition to other quality cuts
the tracks must have a minimum transverse momentum of 120 MeV/$c$
and at least 20 jet chamber hits.
The intersection of the tracks in the $r\phi$~plane is 
considered as a secondary 
vertex candidate if it satisfies the following criteria:
\begin{itemize}
\item the radial distance between primary vertex and the intersection point 
must be greater than 0.5 cm and less than 150 cm. 
For events with at least 6 tracks the primary
vertex is fitted and for events with less than 6 tracks
the beam spot reconstructed from tracks collected from many consecutive 
$\ee$ events during a LEP fill is taken as primary 
vertex~\cite{bib-prim}. 
\item the difference between the radial coordinate of the secondary vertex and
the radial coordinate of the first jet-chamber hit associated with either of
the two tracks has to be less than 10 cm;
\item the radial coordinate of the tracks 
at the point of closest approach to 
the primary vertex has to be greater than 0.2~cm;
\item the angle between the direction of flight from primary to 
secondary vertex and the combined momentum vector
of the two tracks at the intersection point 
has to be less than~5$^{\circ}$.
\end{itemize}
In addition, a fit was performed for track pairs passing all 
these cuts, constraining them to originate from a common vertex.
A correction procedure was used to compensate for the energy
loss of the pions in the inactive material of the detector. 
All secondary vertices satisfying 
$|M(\pi^+\pi^-)-0.4977\mbox{~GeV}/c^2|< 0.02$~GeV/$c^2$ 
are considered to be $\ks$ decay vertices, where
the mass $M$ is calculated assuming that both tracks are pions.
Finally, the residual background is reduced by requiring 
at least 20 ${\rm d}E/{\rm d}x$ hits.
The two tracks are identified as pions if the ${\mathrm d}E/{\mathrm d}x$ 
probability for the pion hypothesis, that is the probability that 
the specific ionisation energy loss in the 
jet chamber $({\rm d}E/{\rm d}x)$ is compatible with that
expected for a pion, exceeds $5\%$.

In Fig.~\ref{fig-mpp} the $\pi^+\pi^-$ invariant mass $M$ 
is shown for all identified secondary vertices in the selected
events before and after applying the  ${ \rm d} E/ {\rm d}x$ 
cuts.
After all cuts the reconstruction efficiency for $\ks\rightarrow\pi^+\pi^-$
decays is about $35.5\%$ and the purity is about $95.5\%$ 
for $\pt(\ks)>1$~GeV/$c$, $|\eta(\ks)|<1.5$ and $10<W<125$~GeV.
\section{Systematic errors}
The following systematic errors, common to the charged
hadron and $\ks$  measurements, are taken
into account:
\begin{itemize}
\item
The correction factors are obtained using PHOJET and PYTHIA, 
separately.
The resulting distributions are averaged to get the final result.
The differences between the two distributions are used to
define the systematic error. 
\item The lower limit on the
trigger efficiency is taken into account by an additional
systematic error of $3\%$ on the cross-section in the range $10<W<30$~GeV.
\item
Systematic errors due to the modelling of the detector resolution
for the measurement of tracks
were found to be negligibly small in comparison to the other errors.
The systematic error due to the uncertainty in the energy scale
of the electromagnetic calorimeter was estimated by varying
the reconstructed ECAL energy in the Monte Carlo by $\pm 5\%$.
\item The limited statistics of the Monte Carlo samples,
especially at large transverse momenta $\pt$, is also included
in the systematic error.
\item The systematic error of the luminosity measurement is
negligible compared to the other systematic errors.
\end{itemize}

The systematic error of the Monte Carlo modelling and of
the ECAL energy scale and, for the low $W$ region, the
error from the trigger efficiency contribute about equally
to the total systematic error.
In the $\Ks$ reconstruction additional
systematic errors were studied by varying the parameters
of the secondary vertex finder and the ${ \rm d} E/ {\rm d}x$ cuts.
The full difference between the results is used to estimate
the contribution to the total systematic error
from the $\Ks$ reconstruction, the Monte Carlo model dependence and
the ECAL energy scale.
The systematic error affecting the $\Ks$ reconstruction 
and the error from comparing the PHOJET and PYTHIA
correction factors are of similar magnitude.
The total systematic error was obtained by adding all
systematic errors in quadrature. The total systematic 
errors are highly correlated from bin to bin.

\section{Results}
\label{sec-results}
\begin{figure}[ht]
   \begin{center}
      \mbox{
          \epsfxsize=15.0cm
          \epsffile{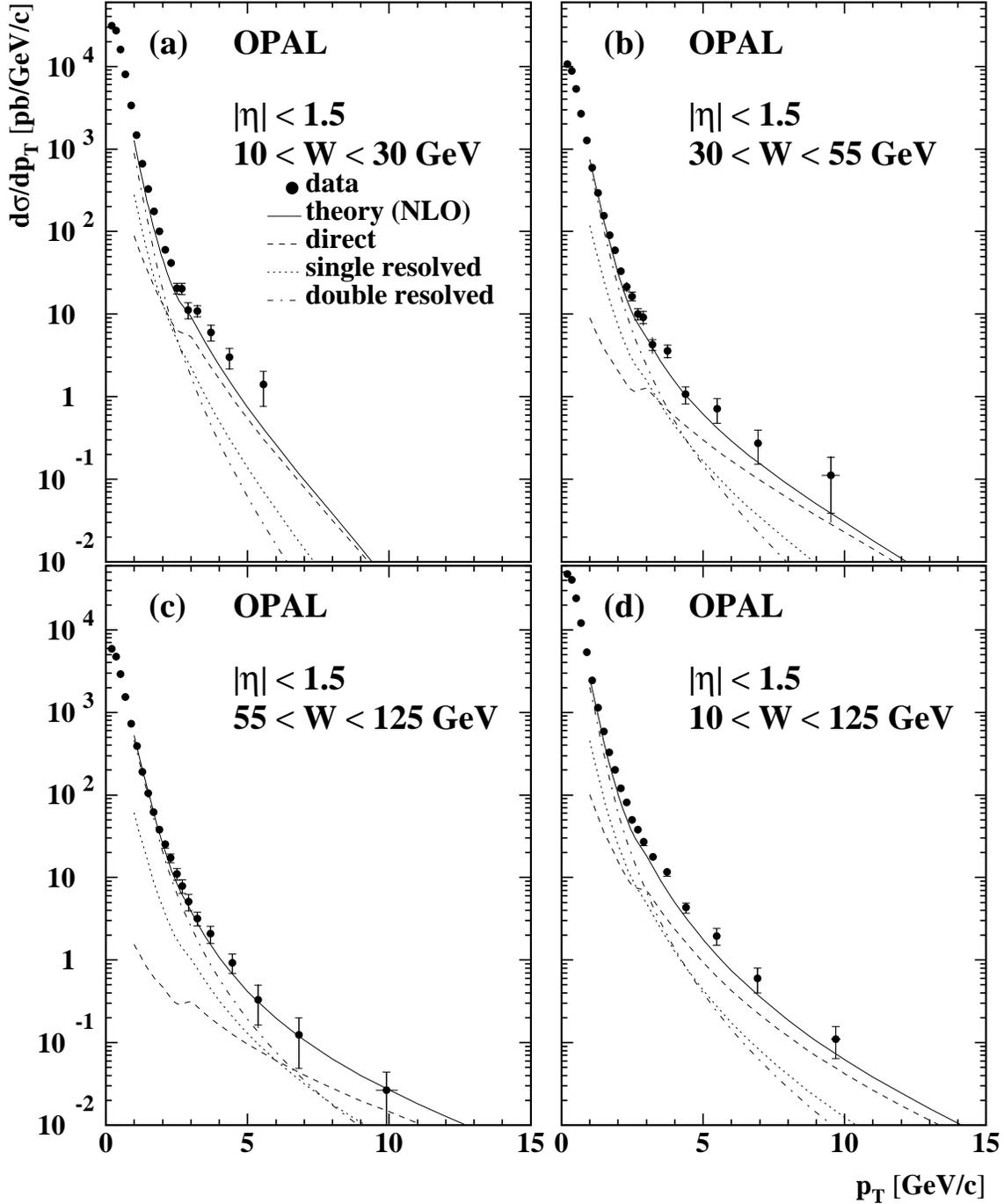}
           }
   \end{center}
\caption{Differential inclusive charged hadron production cross-sections
$\dspt$ for $|\eta|<1.5$ and in the $W$ ranges 
(a) $10<W<30$~GeV;
(b) $30<W<55$~GeV;
(c) $55<W<125$~GeV and 
(d) for all $W$ ($10<W<125$~GeV)
measured at $\sqee=161$ and $172$~GeV.
The data are compared to NLO calculations for $\pt>1$~GeV$/c$
by Binnewies et al. 
together with the separate contributions of double-resolved, single-resolved 
and direct $\gg$ interactions.
The inner error bar shows the statistical error and the outer error bar the
statistical and systematic errors added in quadrature.}
\label{fig-pt1}
\end{figure}

The differential inclusive cross-section $\dspt$
for charged hadrons in the region $|\eta|<1.5$ is 
shown in Fig.~\ref{fig-pt1} for different corrected $W$ ranges
together with the statistical and systematical errors. 
The corrected cross-sections are given in 
Tables~\ref{tab-pt1a} and \ref{tab-pt1b}.

The measured differential cross-sections are compared to NLO
calculations by Binnewies, Kniehl and Kramer~\cite{bib-binnewies}.
The cross-sections are calculated using the QCD partonic cross-sections
to NLO for direct, single- and double-resolved processes.
The hadronic cross-section is a convolution of the Weizs\"acker-Williams 
effective photon distribution, the parton distribution functions and
the fragmentation functions of Ref.~\cite{bib-bkk} which are obtained
from a fit to $\ee$ data from TPC and ALEPH.
The NLO GRV parametrisation of the parton densities of the 
photon~\cite{bib-grv} is
used with $\Lambda^{(5)}_{\overline{\rm MS}}=131$ MeV
and $m_{\rm c}=1.5$~GeV$/c^2$. 
The renormalization and factorization scales
in the calculation are set equal to $\xi\pt$ with $\xi=1$. 
The change in slope around $\pt=3$~GeV/$c$ in the
NLO calculation is due to the charm threshold, below which
the charm distribution in the resolved photon and the charm
fragmentation functions are set to zero.
 
The cross-section calculation was
repeated for the kinematic conditions of the data presented here
at an average $\ee$ centre-of-mass energy
$\sqee=166.5$~GeV and for scattering angles $\theta'<33$~mrad.
For the differential cross-section $\dspt$ 
a minimum $p_{\rm T}$ of 1~GeV/$c$ is required to ensure
the validity of the perturbative QCD calculation.
For the same reason the differential cross-section $\dseta$
is restricted to the region $\pt>1.5$~GeV/$c$.
The scale dependence of the NLO calculation was studied
by setting $\xi=0.5$ and 2. This leads to a variation of the cross-section
of about $30\%$ at $\pt=1$~GeV/$c$ and of about $10\%$ for $\pt>5$~GeV/$c$.
The NLO calculations lie significantly below the data
for $W<30$~GeV for $\dspt$ and $\dseta$. 
The agreement with the data improves in the higher $W$ bins.
The NLO calculation is shown separately for double-resolved, 
single-resolved and direct 
interactions. At large $\pt$ the direct interactions dominate.
It should be noted that these classifications are scale dependent in NLO. 

\begin{figure}[htbp]
   \begin{center}
      \mbox{
          \epsfxsize=16.0cm
          \epsffile{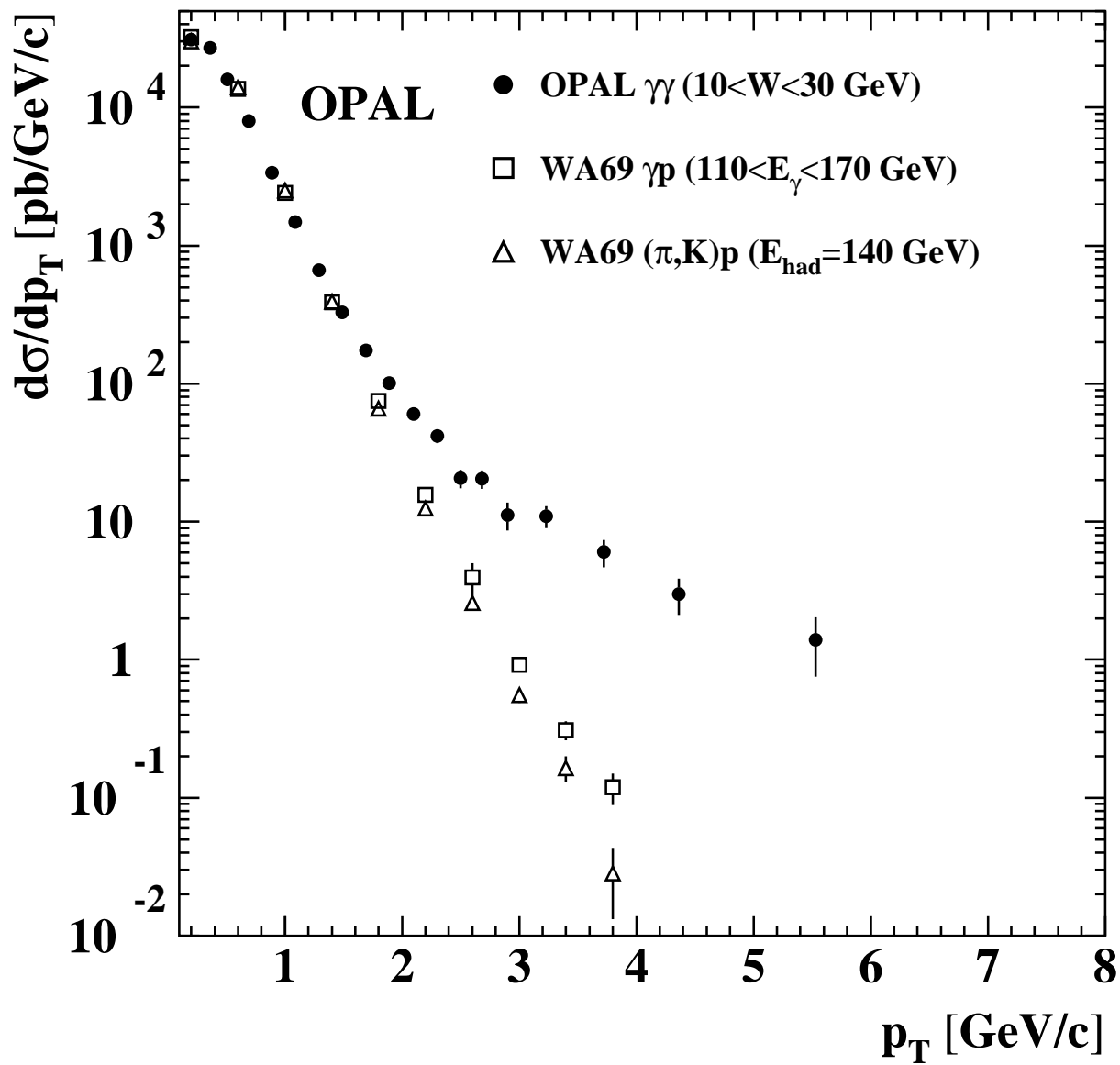}
           }
   \end{center}
\caption{The $\pt$ distribution measured in $\gg$ interactions
in the range $10<W<30$~GeV is compared to the $p_{\rm T}$ distribution
measured in $\gamma$p and hp (h$=\pi,$K) interactions in the experiment 
WA69~\cite{bib-wa69}. The hp and $\gamma$p data are
normalised to the low $p_{\rm T}$ region at $\pt\approx 200$~MeV$/c$.
The cross-section values given on the ordinate
are therefore only valid for the OPAL data.
}
\label{fig-wa69}
\end{figure}

The $p_{\rm T}$ distribution for $10<W<30$~GeV 
is compared in Fig.~\ref{fig-wa69} to $p_{\rm T}$ distributions
in $\gamma$p and hp (h$=\pi,$K) interactions measured by the experiment 
WA69~\cite{bib-wa69}. The hp data are
weighted by WA69 in such a way that they contain $60\%$ $\pi$p and 
$40\%$ Kp data to match the expected mixture of non-strange
and strange quarks in the photon beam of the $\gamma$p data. 
The WA69 data is normalised to the $\gg$ data in the low 
$p_{\rm T}$ region 
at $\pt\approx 200$~MeV/$c$ using the same factor for the hp and the
$\gamma$p data.
The $p_{\rm T}$ distribution of WA69 has been measured
in the Feynman-$x$ range $0.0<x_{\rm F}<1.0$. The hadronic invariant
mass of the hp data is $W=16$~GeV and the average $\langle W\rangle$ 
is of similar
size for the $\gamma$p data. 
In the $\gg$ Monte Carlo the average $\langle W \rangle$ is about 
17~GeV in the range $10<W<30$~GeV,
i.e.~the average values of $W$ in the different data samples are 
approximately the same. Whereas only a small increase is observed
in the $\gamma$p data compared to the  $\pi$p and K$\pi$ data at large $\pt$,
there is a significant increase of the relative rate in the range 
$\pt>2$~GeV/$c$ for $\gg$ interactions due to the
direct process. 
A clear deviation is seen at large $\pt$ from the exponential
fall-off expected for purely hadronic interactions.

\begin{figure}[htbp]
   \begin{center}
      \mbox{
          \epsfxsize=16.0cm
          \epsffile{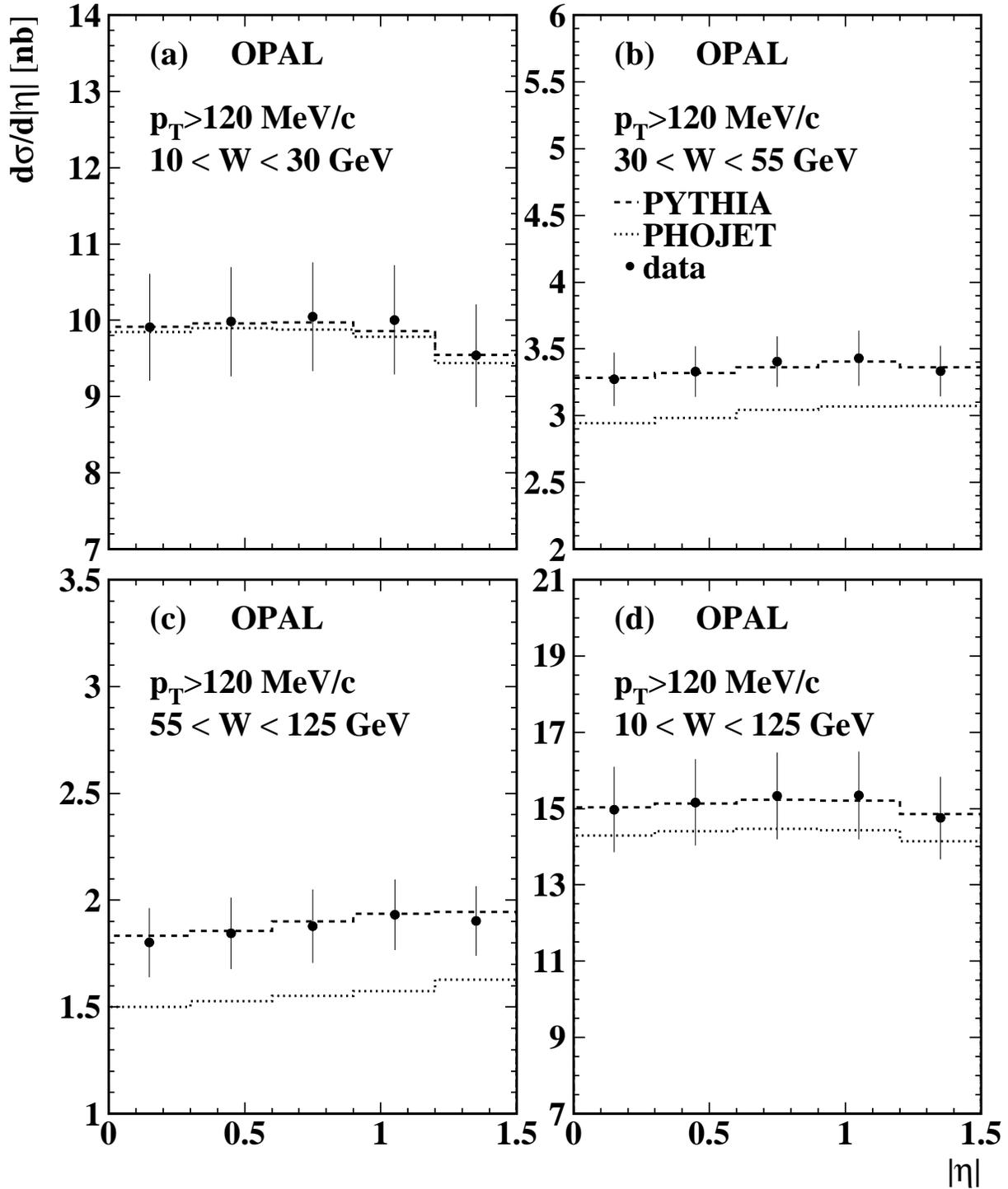}
           }
   \end{center}
\caption{Differential inclusive charged hadron production cross-sections
$\dseta$ for $\pt>120$~MeV$/c$ and in the $W$ ranges 
(a) $10<W<30$~GeV;
(b) $30<W<55$~GeV;
(c) $55<W<~125$~GeV and 
(d) for all $W$ ($10<W<125$~GeV)
measured at $\sqee=161$ and $172$~GeV.
The data are compared to the PHOJET and PYTHIA simulation.
The Monte Carlo distributions are 
plotted as histograms using the same bin width as for the data.
The statistical error is smaller than the symbol size. The error bars show
the statistical and systematic errors added in quadrature.}
\label{fig-eta1}
\end{figure}

The differential cross-section $\dseta$ is compared to
the predictions of the Monte Carlo generators PHOJET 1.10
and PYTHIA 5.722 in Fig.~\ref{fig-eta1} taking into account
the anti-tagging condition $\theta'<33$~mrad.
In PHOJET the $Q^2$ suppression of the total $\gg$ cross-section
is parametrised using Generalised Vector Meson Dominance (GVMD) and
a model for the change of soft hadron production and diffraction with
increasing photon virtuality $Q^2$ is also included.  
The photon-photon mode of PYTHIA only simulates the interactions
of real photons with $Q^2=0$. The virtuality of the photons defined
by $Q^2$ enters only through the equivalent photon
approximation in the generation of the photon 
energy spectrum, but the electrons are scattered at zero angle.
This model is not expected to be correct for larger values of $Q^2$.
We have therefore simulated events with $Q^2<1$~GeV$^2$ with the
photon-photon mode of PYTHIA and events with $Q^2>1$~GeV$^2$ and
$\theta'<33$~mrad with the electron-photon mode of PYTHIA.

\begin{figure}[htbp]
   \begin{center}
      \mbox{
          \epsfxsize=16.0cm
          \epsffile{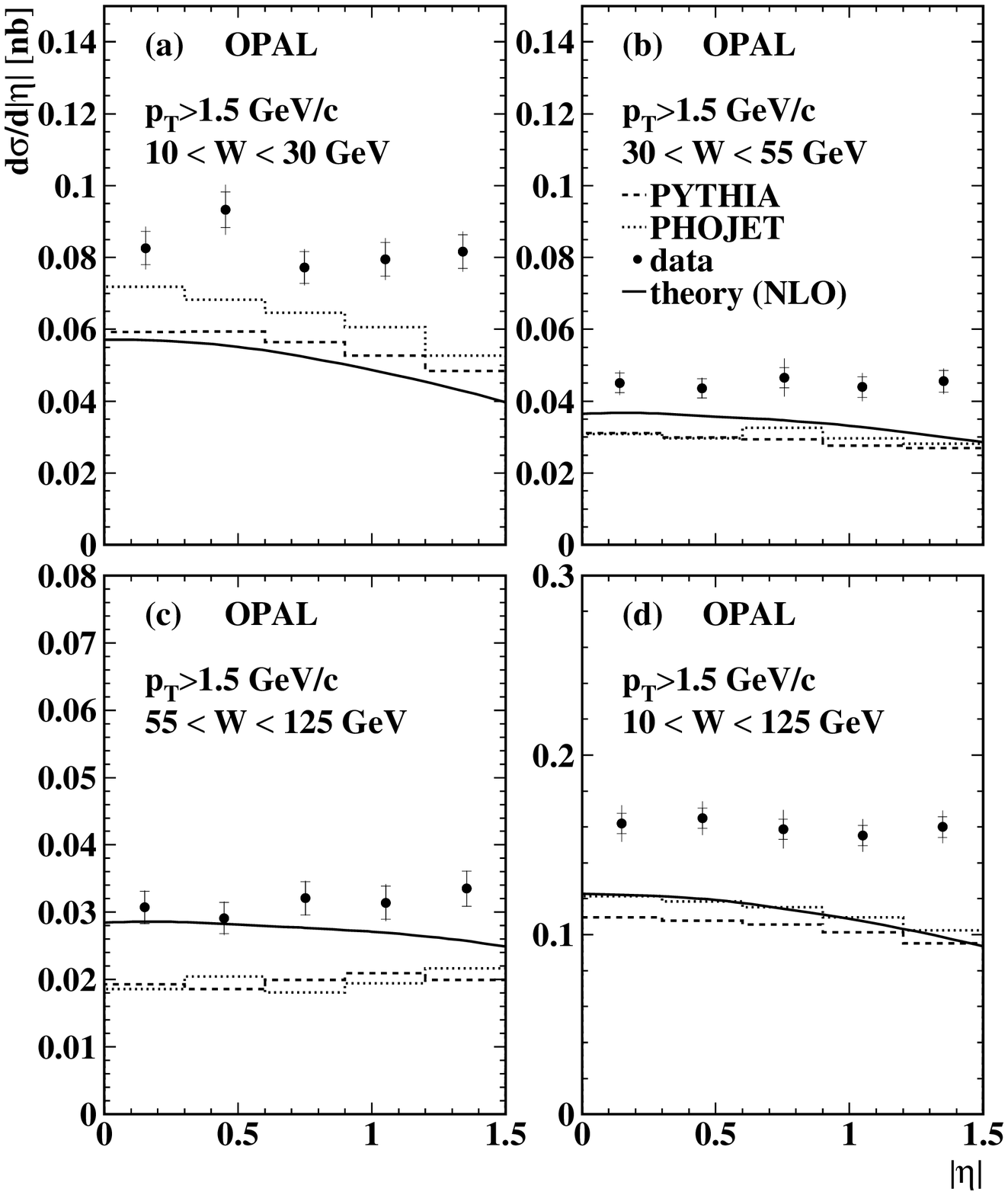}
           }
   \end{center} 
\caption{Differential inclusive charged hadron production cross-sections
$\dseta$ for $\pt>1.5$~GeV$/c$ and in the $W$ ranges 
(a) $10<W<30$~GeV;
(b) $30<W<55$~GeV;
(c) $55<~W<125$~GeV and 
(d) for all $W$ ($10<W<125$~GeV)
measured at $\sqee=161$ and $172$~GeV.
The data are compared to the PHOJET and PYTHIA simulation
and to NLO calculations.
The Monte Carlo distributions are 
plotted as histograms using the same bin width as for the data.
The inner error bar shows the statistical error and the outer error bar the
statistical and systematic errors  added in quadrature.}
\label{fig-eta2}
\end{figure}

The differential cross-section $\dseta$ shown in Fig.~\ref{fig-eta1}
is nearly independent of $|\eta|$ in the measured range.
The $|\eta|$ distribution is reasonably well described by 
PYTHIA and PHOJET for $\pt>120$~MeV$/c$, 
apart from the high $W$ region where PHOJET appears to be below the data.
For transverse momenta $\pt>1.5$~GeV$/c$ (Fig.~\ref{fig-eta2})
both Monte Carlo models underestimate the data significantly.
The same behaviour is observed for the NLO calculation at low $W$, but
the agreement of the NLO calculation with the data improves in the 
high $W$ bins.
The corrected cross-sections are given in 
Tables~\ref{tab-eta2} and \ref{tab-eta1}.

\begin{figure}[htbp]
\begin{tabular}{cc}
\epsfig{file=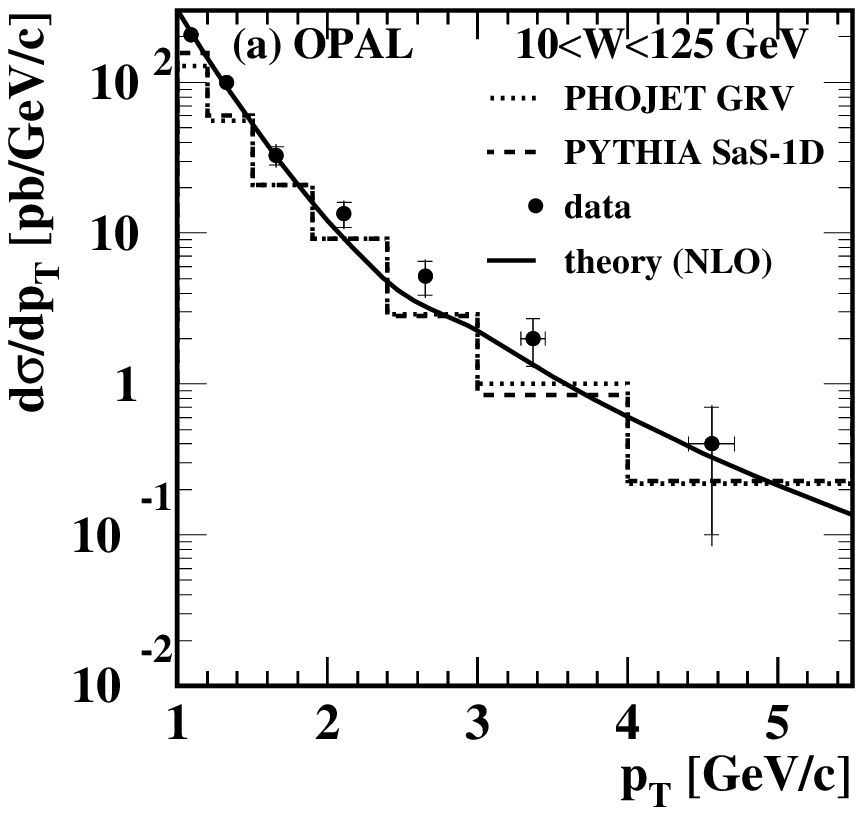,width=0.463\textwidth,height=0.463\textwidth}
 &
\epsfig{file=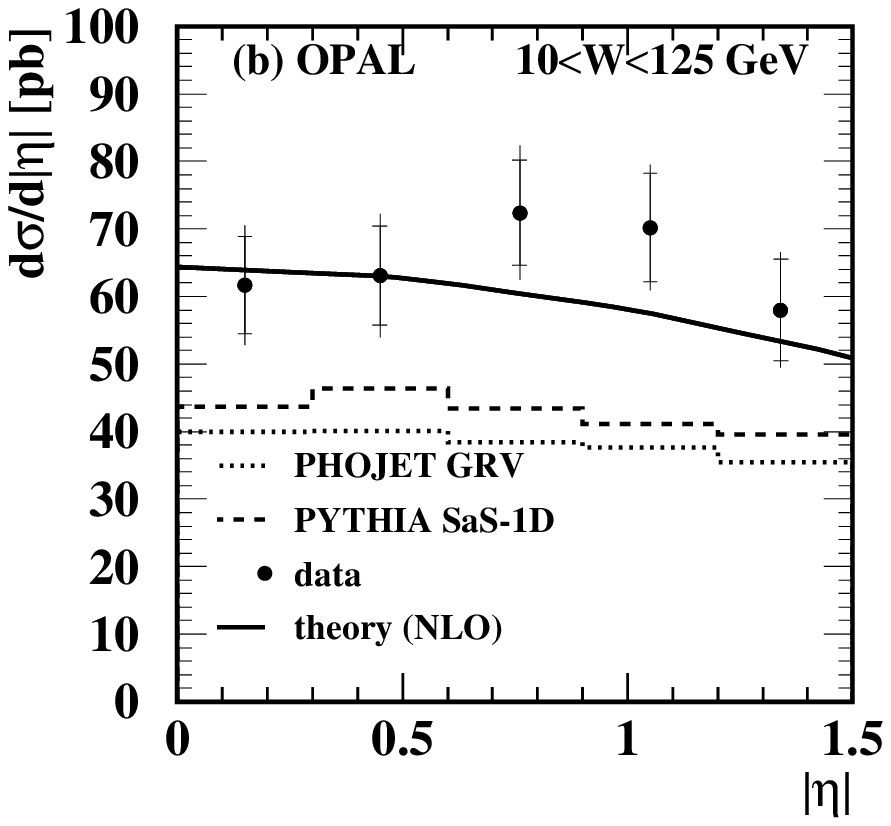,width=0.463\textwidth,height=0.463\textwidth} 
\end{tabular}
\caption{Differential inclusive $\ks$ production cross-sections
(a) $\dspt$ and (b) $\dseta$
for $\pt(\ks)>1$~GeV$/c$ and $|\eta(\ks)|<1.5$ 
in the $W$ range $10<W<125$~GeV. The data are
compared to the PHOJET and PYTHIA simulation
and to NLO calculations.
The data were taken at $\sqee=161$ and $172$~GeV.
The Monte Carlo distributions are 
plotted as histograms using the same bin width as for the data.
The inner error bar shows the statistical error and the outer error bar the
statistical and systematic errors added in quadrature. 
}
\label{fig-ks1}
\end{figure}

\begin{figure}[htbp]
\begin{tabular}{cc}
\epsfig{file=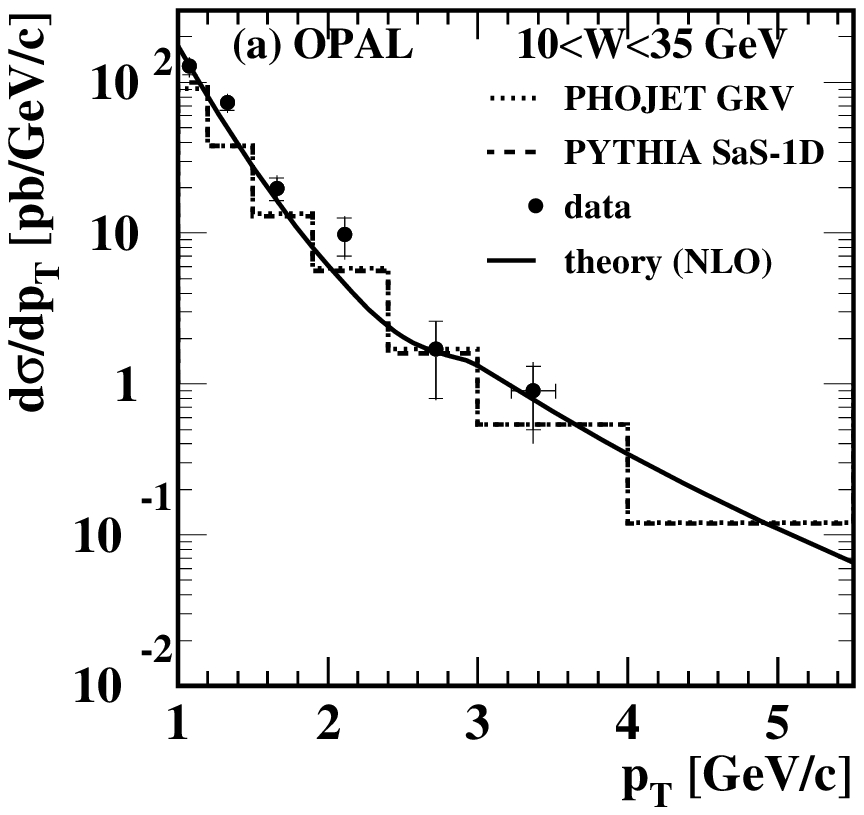,width=0.463\textwidth,height=0.463\textwidth}
 &
\epsfig{file=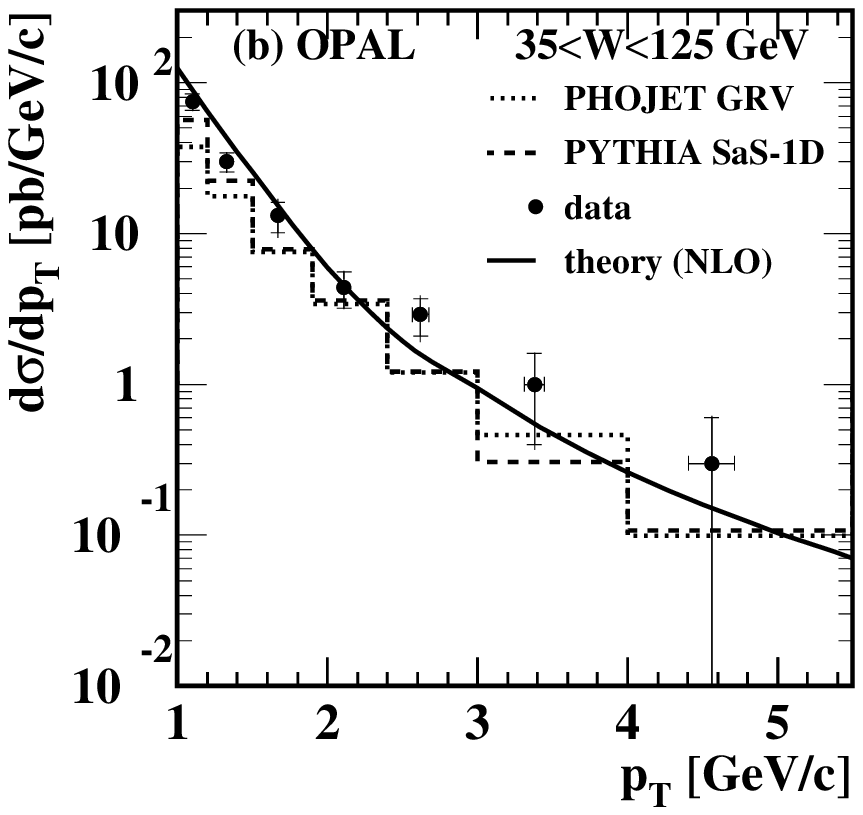,width=0.463\textwidth,height=0.463\textwidth} 
\end{tabular}
\caption{Differential inclusive $\ks$ production cross-sections
$\dspt$ for $\pt(\ks)>1$~GeV$/c$ and $|\eta(\ks)|<1.5$ 
in the $W$ ranges (a) $10<W<35$~GeV and (b) $35<W<125$~GeV.
The data are
compared to the PHOJET and PYTHIA simulation
and to NLO calculations.
The data were taken at $\sqee=161$ and $172$~GeV.
The Monte Carlo distributions are 
plotted as histograms using the same bin width as for the data.
The inner error bar shows the statistical error and the outer error bar the
statistical and systematic errors added in quadrature. 
}
\label{fig-ks2}
\end{figure}

The differential inclusive cross-sections $\dspt$
and $\dseta$ have been measured for $\ks$ mesons
with $\pt(\ks)>1$~GeV$/c$ and $|\eta(\ks)|<1.5$. 
The $\pt$ and $\eta$ dependent cross-sections
are presented in the $W$ range $10~<~W~<~125$~GeV
(Fig.~\ref{fig-ks1} and Tables~\ref{tab-ks1}--\ref{tab-ks2}).
In addition the $\pt$ distribution is shown for
two separate $W$ ranges (Fig.~\ref{fig-ks2} and Table~\ref{tab-ks3}).
The results are compared to PHOJET and PYTHIA.
Based on the PYTHIA simulation using SaS-1D, about half of the 
$\ks$ are expected to be produced from charm quarks
at large $\pt$ where direct processes are dominant.
Both Monte Carlo models significantly underestimate the $\ks$ production
cross-section in the low $\pt$ region where most $\ks$ are
expected to originate from primary strange quarks. 
The distributions are reasonably well 
described by the NLO calculations
which use the $\ks$ fragmentation function fitted
to MARK II~\cite{bib-markk0} and ALEPH~\cite{bib-alephk0}
data in Ref.~\cite{bib-fk0}.
The change in slope between $\pt=2$ and $\pt=3$~GeV/$c$ in the
NLO calculation is again due to the charm threshold.
The variation of the calculated cross-section
as a function of $\pt$ for different choices of scales, $\xi=0.5$ and $2$,
is largest around the charm threshold,
about $30-40\%$, and $10-20\%$ elsewhere.

\section{Conclusions}
We present measurements of differential cross-sections
as a function of transverse momentum and pseudorapidity
for charged hadrons and $\Ks$ mesons produced in photon-photon
collisions at LEP. The data were taken at $\ee$ centre-of-mass
energies of 161 and 172~GeV. 

The differential cross-section $\dspt$ for charged hadrons 
is compared to NLO calculations.
In the range $10<W<30$~GeV more charged hadrons are found 
at large $\pt$ than predicted. 
Good agreement between the NLO calculation and the data is
found in the highest $W$ range, $55<W<125$~GeV. 
The Monte Carlo models PYTHIA and PHOJET both underestimate 
the cross-section for tracks with $\pt>1.5$~GeV and $|\eta|<1.5$
in all $W$ ranges.
The shape of the differential cross-section $\dseta$ is
well reproduced by the NLO calculations and the Monte Carlo models.
A comparison of the $\pt$ distributions of the $\gg$ data
to $\pt$ distributions measured in $\gamma$p and ($\pi$,K)p processes
at similar invariant masses shows the relative increase of hard interactions
in $\gg$ processes due to the direct component.
The transverse momentum and pseudorapidity distributions
of the $\ks$ mesons are reasonably well reproduced by the NLO calculations,
but they are significantly underestimated by the Monte Carlo models
PHOJET and PYTHIA.

\medskip
\bigskip\bigskip\bigskip
\appendix
\par
\section*{Acknowledgements}
\par
We thank Janko Binnewies, Bernd Kniehl and Gustav Kramer for providing 
the NLO calculations and for many useful discussions.\\
We particularly wish to thank the SL Division for the efficient operation
of the LEP accelerator at all energies
 and for their continuing close cooperation with
our experimental group.  We thank our colleagues from CEA, DAPNIA/SPP,
CE-Saclay for their efforts over the years on the time-of-flight and trigger
systems which we continue to use.  In addition to the support staff at our own
institutions we are pleased to acknowledge the  \\
Department of Energy, USA, \\
National Science Foundation, USA, \\
Particle Physics and Astronomy Research Council, UK, \\
Natural Sciences and Engineering Research Council, Canada, \\
Israel Science Foundation, administered by the Israel
Academy of Science and Humanities, \\
Minerva Gesellschaft, \\
Benoziyo Center for High Energy Physics,\\
Japanese Ministry of Education, Science and Culture (the
Monbusho) and a grant under the Monbusho International
Science Research Program,\\
German Israeli Bi-national Science Foundation (GIF), \\
Bundesministerium f\"ur Bildung, Wissenschaft,
Forschung und Technologie, Germany, \\
National Research Council of Canada, \\
Research Corporation, USA,\\
Hungarian Foundation for Scientific Research, OTKA T-016660, 
T023793 and OTKA F-023259.\\

\newpage

\begin{table}[htbp]
\begin{center} 
\begin{tabular}{|c||l|l||l|l|} \hline
 & \multicolumn{2}{c||}{$10<W<30$~GeV} &  \multicolumn{2}{c|}{$30<W<55$~GeV} \\ \cline{2-5}
$\pt$ [GeV/$c$] & $\langle \pt \rangle$ [GeV/$c$] &
\multicolumn{1}{c||}{$\dspt$~[pb/GeV/$c$]} & $\langle \pt \rangle$ [GeV/$c$] &
\multicolumn{1}{c|}{$\dspt$~[pb/GeV/$c$]} \\ \hline

0.12-0.28 & 0.20 & (3.11$\pm$0.01$\pm$0.24)$\times 10^{4}$ & 
0.20 & (1.06$\pm$0.01$\pm$0.08)$\times 10^{4}$ \\
0.28-0.44 & 0.36 & (2.70$\pm$0.01$\pm$0.20)$\times 10^{4}$ & 
0.35 & (8.80$\pm$0.06$\pm$0.58)$\times 10^{3}$ \\
0.44-0.60 & 0.51 & (1.60$\pm$0.01$\pm$0.10)$\times 10^{4}$ & 
0.51 & (5.31$\pm$0.04$\pm$0.32)$\times 10^{3}$ \\
0.60-0.80 & 0.69 & (8.00$\pm$0.06$\pm$0.46)$\times 10^{3}$ & 
0.69 & (2.67$\pm$0.03$\pm$0.14)$\times 10^{3}$ \\
0.80-1.00 & 0.89 & (3.38$\pm$0.04$\pm$0.17)$\times 10^{3}$ & 
0.89 & (1.26$\pm$0.02$\pm$0.06)$\times 10^{3}$ \\
1.00-1.20 & 1.09 & (1.48$\pm$0.02$\pm$0.08)$\times 10^{3}$ & 
1.09 & (5.95$\pm$0.12$\pm$0.22)$\times 10^{2}$ \\
1.20-1.40 & 1.29 & (6.64$\pm$0.16$\pm$0.40)$\times 10^{2}$ & 
1.29 & (2.92$\pm$0.09$\pm$0.12)$\times 10^{2}$ \\
1.40-1.60 & 1.49 & (3.29$\pm$0.11$\pm$0.21)$\times 10^{2}$ & 
1.49 & (1.55$\pm$0.07$\pm$0.06)$\times 10^{2}$ \\
1.60-1.80 & 1.69 & (1.75$\pm$0.08$\pm$0.11)$\times 10^{2}$ & 
1.69 & (9.01$\pm$0.49$\pm$0.55)$\times 10^{1}$ \\
1.80-2.00 & 1.89 & (1.00$\pm$0.06$\pm$0.07)$\times 10^{2}$ & 
1.89 & (5.96$\pm$0.39$\pm$0.38)$\times 10^{1}$ \\
2.00-2.20 & 2.10 & (6.04$\pm$0.48$\pm$0.37)$\times 10^{1}$ & 
2.09 & (3.29$\pm$0.29$\pm$0.13)$\times 10^{1}$ \\
2.20-2.40 & 2.30 & (4.18$\pm$0.39$\pm$0.27)$\times 10^{1}$ & 
2.29 & (2.15$\pm$0.23$\pm$0.20)$\times 10^{1}$ \\
2.40-2.60 & 2.50 & (2.06$\pm$0.30$\pm$0.08)$\times 10^{1}$ & 
2.50 & (1.64$\pm$0.20$\pm$0.07)$\times 10^{1}$ \\
2.60-2.80 & 2.68 & (2.04$\pm$0.31$\pm$0.07)$\times 10^{1}$ & 
2.70 & (1.01$\pm$0.16$\pm$0.08)$\times 10^{1}$ \\
2.80-3.00 & 2.90 & (1.12$\pm$0.25$\pm$0.05)$\times 10^{1}$ & 
2.88 & (9.18$\pm$1.55$\pm$1.10) \\
3.00-3.50 & 3.22 & (1.09$\pm$0.16$\pm$0.11)$\times 10^{1}$ & 
3.21 & (4.26$\pm$0.65$\pm$0.68) \\
3.50-4.00 & 3.71$\pm$0.01 & (6.03$\pm$1.35$\pm$0.21) & 
3.74$\pm$0.01 & (3.58$\pm$0.62$\pm$0.32) \\
4.00-5.00 & 4.36$\pm$0.01 & (2.99$\pm$0.82$\pm$0.33) & 
4.38$\pm$0.02 & (1.07$\pm$0.25$\pm$0.04) \\
5.00-6.00 & 5.55$\pm$0.03 & (1.40$\pm$0.64$\pm$0.07) & 
5.50$\pm$0.04 & (7.12$\pm$2.38$\pm$0.44)$\times 10^{-1}$ \\
6.00-8.00 & -- & -- & 
6.94$\pm$0.07 & (2.74$\pm$1.22$\pm$0.15)$\times 10^{-1}$ \\
8.00-15.0 & -- & -- & 
9.51$\pm$0.32 & (1.11$\pm$0.73$\pm$0.36)$\times 10^{-1}$ \\
 \hline
\end{tabular}
\caption{Differential inclusive charged hadron production cross-sections
$\dspt$ for $|\eta|<1.5$ and in the $W$ ranges $10<W<30$~GeV and 
$30<W<55$~GeV. 
The first error is statistical and the second is systematic.
No value is given if the error on $\langle \pt \rangle$ is less than 0.01.}
\label{tab-pt1a}
\end{center}
\end{table}

\newpage
\begin{table}[htbp]
\begin{center} 
\begin{tabular}{|c||l|l||l|l|} \hline
 & \multicolumn{2}{c||}{$55<W<125$~GeV} & \multicolumn{2}{c|}{$10<W<125$~GeV} 
 \\ \cline{2-5}
$\pt$ [GeV/$c$] & $\langle \pt \rangle$ [GeV/$c$] & 
\multicolumn{1}{c||}{$\dspt$~[pb/GeV/$c$]} & $\langle \pt \rangle$ [GeV/$c$] &
\multicolumn{1}{c|}{$\dspt$~[pb/GeV/$c$]} \\ \hline
0.12-0.28 & \pz0.20 & (5.85$\pm$0.05$\pm$0.58)$\times 10^{3}$ & 
0.20 & (4.78$\pm$0.02$\pm$0.38)$\times 10^{4}$ \\
0.28-0.44 &\pz0.35 & (4.75$\pm$0.04$\pm$0.46)$\times 10^{3}$ & 
0.36 & (4.05$\pm$0.01$\pm$0.31)$\times 10^{4}$ \\
0.44-0.60 &\pz0.51 & (2.92$\pm$0.03$\pm$0.28)$\times 10^{3}$ & 
0.51 & (2.42$\pm$0.01$\pm$0.17)$\times 10^{4}$ \\
0.60-0.80 &\pz0.69 & (1.53$\pm$0.02$\pm$0.12)$\times 10^{3}$ & 
0.69 & (1.21$\pm$0.01$\pm$0.07)$\times 10^{4}$ \\
0.80-1.00 &\pz0.89 & (7.25$\pm$0.15$\pm$0.47)$\times 10^{2}$ & 
0.89 & (5.33$\pm$0.04$\pm$0.27)$\times 10^{3}$ \\
 1.00-1.20 &\pz1.09 & (3.89$\pm$0.10$\pm$0.20)$\times 10^{2}$ & 
 1.09 & (2.45$\pm$0.03$\pm$0.12)$\times 10^{3}$ \\
 1.20-1.40 &\pz1.29 & (1.91$\pm$0.07$\pm$0.09)$\times 10^{2}$ & 
 1.29 & (1.14$\pm$0.02$\pm$0.06)$\times 10^{3}$ \\
 1.40-1.60 &\pz1.49 & (1.04$\pm$0.06$\pm$0.04)$\times 10^{2}$ & 
 1.49 & (5.87$\pm$0.13$\pm$0.31)$\times 10^{2}$ \\
 1.60-1.80 &\pz1.69 & (6.20$\pm$0.41$\pm$0.18)$\times 10^{1}$ & 
 1.69 & (3.28$\pm$0.10$\pm$0.19)$\times 10^{2}$ \\
 1.80-2.00 &\pz1.89 & (3.78$\pm$0.33$\pm$0.20)$\times 10^{1}$ & 
 1.89 & (2.00$\pm$0.08$\pm$0.11)$\times 10^{2}$ \\
 2.00-2.20 &\pz2.09 & (2.53$\pm$0.27$\pm$0.06)$\times 10^{1}$ & 
 2.09 & (1.20$\pm$0.06$\pm$0.06)$\times 10^{2}$ \\
 2.20-2.40 &\pz2.28 & (1.72$\pm$0.22$\pm$0.04)$\times 10^{1}$ & 
 2.29 & (8.10$\pm$0.49$\pm$0.57)$\times 10^{1}$ \\
 2.40-2.60 &\pz2.50 & (1.10$\pm$0.18$\pm$0.02)$\times 10^{1}$ & 
 2.50 & (5.00$\pm$0.38$\pm$0.20)$\times 10^{1}$ \\
 2.60-2.80 &\pz2.69 & (7.89$\pm$1.49$\pm$0.23)  &
 2.69 & (3.81$\pm$0.34$\pm$0.18)$\times 10^{1}$ \\
 2.80-3.00 &\pz2.92 & (5.09$\pm$1.15$\pm$0.34)  & 
 2.90 & (2.70$\pm$0.29$\pm$0.11)$\times 10^{1}$ \\
 3.00-3.50 &\pz3.23 & (3.18$\pm$0.59$\pm$0.04)  & 
 3.22 & (1.77$\pm$0.16$\pm$0.07)$\times 10^{1}$ \\
 3.50-4.00 &\pz3.70$\pm$0.01 & (2.07$\pm$0.48$\pm$0.10)  & 
 3.72 & (1.16$\pm$0.13$\pm$0.04)$\times 10^{1}$ \\
 4.00-5.00 &\pz4.46$\pm$0.02 & (9.30$\pm$2.44$\pm$0.58)$\times 10^{-1}$ & 
 4.40$\pm$0.01 & (4.32$\pm$0.61$\pm$0.19)  \\
 5.00-6.00 &\pz5.38$\pm$0.04 & (3.30$\pm$1.67$\pm$0.64)$\times 10^{-1}$ & 
 5.48$\pm$0.01 & (1.95$\pm$0.45$\pm$0.09)  \\
 6.00-8.00 &\pz6.81$\pm$0.09 & (1.24$\pm$0.76$\pm$0.05)$\times 10^{-1}$ & 
 6.92$\pm$0.03 & (5.97$\pm$2.01$\pm$0.40)$\times 10^{-1}$ \\
 8.00-15.0 & \pz 9.91$\pm$0.34 & (2.64$\pm$1.76$\pm$0.23)$\times 10^{-2}$ & 
 9.69$\pm$0.18 & (1.10$\pm$0.46$\pm$0.07)$\times 10^{-1}$ \\
 \hline
\end{tabular}
\caption{Differential inclusive charged hadron production cross-sections
$\dspt$ for $|\eta|<1.5$ and in the $W$ range  
$55<W<125$~GeV and for all $W$ ($10<W<125$~GeV).
The first error is the statistical error and
the second error is the systematic error.
No value is given if the error on $\langle \pt \rangle$ is less than 0.01.}
\label{tab-pt1b}
\end{center}
\end{table}

\begin{table}[htbp]
\begin{center} 
\begin{tabular}{|c||l|l||l|l|} \hline
  & \multicolumn{2}{|c||}{$10<W<30$ GeV} & \multicolumn{2}{c|}{$30<W<55$ GeV} \\ \cline{2-5}
$|\eta|$ & $\langle |\eta| \rangle$ &\multicolumn{1}{c||}{$\dseta$~[nb]} &
$\langle  |\eta | \rangle$ &\multicolumn{1}{c|}{$\dseta$~[nb]} \\ \hline
 0.00-0.30 & 0.15 & \pz9.91$\pm$0.06$\pm$0.70 & 
 0.15 & 3.27$\pm$0.03$\pm$0.20 \\
 0.30-0.60 & 0.45 & \pz9.98$\pm$0.06$\pm$0.71 & 
 0.45 & 3.33$\pm$0.03$\pm$0.19 \\
 0.60-0.90 & 0.75 &   10.05$\pm$0.06$\pm$0.71 & 
 0.75 & 3.41$\pm$0.03$\pm$0.19 \\
 0.90-1.20 & 1.05 &   10.01$\pm$0.06$\pm$0.71 & 
 1.05 & 3.43$\pm$0.03$\pm$0.20 \\
 1.20-1.50 & 1.35 & \pz9.54$\pm$0.06$\pm$0.67 & 
 1.35 & 3.33$\pm$0.03$\pm$0.19 \\
 \hline
 & \multicolumn{2}{|c||}{$55<W<125$ GeV} & \multicolumn{2}{c|}{$10<W<125$ GeV} \\ \cline{2-5}
$|\eta|$  & $\langle |\eta| \rangle$ &\multicolumn{1}{c||}{$\dseta$~[nb]} &
$\langle  |\eta | \rangle$ &\multicolumn{1}{c|}{$\dseta$~[nb]} \\ \hline
 0.00-0.30 & 0.15 & 1.80$\pm$0.02$\pm$0.16 & 
 0.15 & 14.97$\pm$0.06$\pm$1.12 \\
 0.30-0.60 & 0.45 & 1.85$\pm$0.02$\pm$0.17 & 
 0.45 & 15.16$\pm$0.07$\pm$1.13 \\
 0.60-0.90 & 0.75 & 1.88$\pm$0.02$\pm$0.17 & 
 0.75 & 15.33$\pm$0.07$\pm$1.14 \\
 0.90-1.20 & 1.05 & 1.93$\pm$0.02$\pm$0.16 & 
 1.05 & 15.35$\pm$0.07$\pm$1.15 \\
 1.20-1.50 & 1.35 & 1.90$\pm$0.02$\pm$0.16 & 
 1.35 & 14.75$\pm$0.06$\pm$1.08 \\
 \hline
\end{tabular}
\caption{Differential inclusive charged hadron production cross-sections
$\dseta$ for $\pt>120$~MeV$/c$ and in the $W$ ranges 
$10<W<30$~GeV, $30<W<55$~GeV,
$55<W<125$~GeV and for all $W$ ($10<W<125$~GeV).
The first error is the statistical error and
the second error is the systematic error.}
\label{tab-eta2}
\end{center}
\end{table} 

\begin{table}[htbp]
\begin{center} 
\begin{tabular}{|c||l|l||l|l|} \hline
 & \multicolumn{2}{|c||}{$10<W<30$ GeV} & \multicolumn{2}{c|}{$30<W<55$ GeV} \\ \cline{2-5}
$|\eta|$ & $\langle |\eta| \rangle$ &\multicolumn{1}{c||}{$\dseta$~[nb]} &
$\langle  |\eta | \rangle$ &\multicolumn{1}{c|}{$\dseta$~[nb]} \\ \hline
 0.00-0.30 & 0.15 & (8.26$\pm$0.47$\pm$0.37)$\times 10^{-2}$ & 
 0.14 & (4.51$\pm$0.27$\pm$0.20)$\times 10^{-2}$ \\
 0.30-0.60 & 0.45 & (9.33$\pm$0.49$\pm$0.49)$\times 10^{-2}$ & 
 0.45 & (4.36$\pm$0.27$\pm$0.09)$\times 10^{-2}$ \\
 0.60-0.90 & 0.75 & (7.72$\pm$0.45$\pm$0.24)$\times 10^{-2}$ & 
 0.76 & (4.66$\pm$0.28$\pm$0.44)$\times 10^{-2}$ \\
 0.90-1.20 & 1.05 & (7.95$\pm$0.47$\pm$0.35)$\times 10^{-2}$ & 
 1.05 & (4.39$\pm$0.29$\pm$0.27)$\times 10^{-2}$ \\
 1.20-1.50 & 1.34 & (8.17$\pm$0.47$\pm$0.31)$\times 10^{-2}$ & 
 1.35 & (4.56$\pm$0.30$\pm$0.19)$\times 10^{-2}$ \\
 \hline
  & \multicolumn{2}{|c||}{$55<W<125$ GeV} & \multicolumn{2}{c|}{$10<W<125$ GeV} \\ \cline{2-5}
$|\eta|$  & $\langle |\eta| \rangle$ & \multicolumn{1}{c||}{$\dseta$~[nb]} &
$\langle  |\eta | \rangle$ &\multicolumn{1}{c|}{$\dseta$~[nb]} \\ \hline
 0.00-0.30 & 0.15 & (3.07$\pm$0.24$\pm$0.07)$\times 10^{-2}$ & 
 0.15 & (1.62$\pm$0.06$\pm$0.09)$\times 10^{-1}$ \\
 0.30-0.60 & 0.45 & (2.91$\pm$0.23$\pm$0.09)$\times 10^{-2}$ & 
 0.45 & (1.65$\pm$0.06$\pm$0.08)$\times 10^{-1}$ \\
 0.60-0.90 & 0.75 & (3.21$\pm$0.25$\pm$0.07)$\times 10^{-2}$ & 
 0.75 & (1.59$\pm$0.06$\pm$0.09)$\times 10^{-1}$ \\
 0.90-1.20 & 1.05 & (3.14$\pm$0.25$\pm$0.12)$\times 10^{-2}$ & 
 1.05 & (1.55$\pm$0.06$\pm$0.07)$\times 10^{-1}$ \\
 1.20-1.50 & 1.35 & (3.35$\pm$0.26$\pm$0.07)$\times 10^{-2}$ & 
 1.35 & (1.60$\pm$0.06$\pm$0.07)$\times 10^{-1}$ \\
 \hline
\end{tabular}
\caption{Differential inclusive charged hadron production cross-sections
$\dseta$ for $\pt>1.5$~GeV$/c$ and in the $W$ ranges 
$10<W<30$~GeV, $30<W<55$~GeV,
$55<W<125$~GeV and for all $W$ ($10<W<125$~GeV).
The first error is the statistical error and
the second error is the systematic error.}
\label{tab-eta1}
\end{center}
\end{table} 

\begin{table}[htbp]
\begin{center}
\begin{tabular}{|c||l|c|}
\hline 
 & \multicolumn{2}{c||}{$10<W<125$~GeV}
\\ \cline{2-3}
 $\pt$ [GeV$/c$] & $\langle \pt \rangle$ [GeV$/c$] & 
d$\sigma$/d$\pt$ [pb/GeV$/c$] 
\\ \hline 
1.0--1.2 & 1.09 $\pm$0.01 & 206.2    $\pm$ 17.4   $\pm$ 16.1\\
1.2--1.5 & 1.33 $\pm$0.01 & 100.2    $\pm$ \pz8.9 $\pm$ \pz8.5 \\
1.5--1.9 & 1.66 $\pm$0.01 & \pz32.9  $\pm$ \pz4.5 $\pm$ \pz3.7 \\
1.9--2.4 & 2.11 $\pm$0.02 & \pz13.5  $\pm$ \pz2.6 $\pm$ \pz1.3 \\
2.4--3.0 & 2.65 $\pm$0.04 & \pzz5.2  $\pm$ \pz1.3 $\pm$ \pz0.7 \\
3.0--4.0 & 3.37 $\pm$0.08 & \pzz 2.0 $\pm$ \pz0.7 $\pm$ \pz0.2 \\
4.0--5.5 & 4.56 $\pm$0.15 & \pzz 0.4 $\pm$ \pz0.3 $\pm$ \pz0.1 \\ \hline
\end{tabular}
\caption{Differential inclusive $\ks$ production cross-sections
$\dspt$ for $\pt(\ks)>1$~GeV$/c$ and $|\eta(\ks)|<1.5$ 
in the $W$ range $10<W<125$~GeV.
The first error is the statistical error and
the second error is the systematic error.} 
\label{tab-ks1}
\end{center}
\end{table}

\begin{table}[htbp]
\begin{center}
\begin{tabular}{|c||c|c|}
\hline  
$|\eta|$  &  $\langle |\eta| \rangle$ & d$\sigma$/d$|\eta|$ [pb]  \\ \hline 
 0.0--0.3 & 0.15 $\pm$ 0.01 & 61.7  $\pm$  7.2 $\pm$ 5.2 \\
 0.3--0.6 & 0.45 $\pm$ 0.01 & 63.1  $\pm$  7.3 $\pm$ 5.4 \\
 0.6--0.9 & 0.76 $\pm$ 0.01 & 72.4  $\pm$  7.8 $\pm$ 6.1 \\
 0.9--1.2 & 1.05 $\pm$ 0.01 & 70.2  $\pm$  8.0 $\pm$ 4.7 \\
 1.2--1.5 & 1.34 $\pm$ 0.01 & 58.0  $\pm$  7.5 $\pm$ 4.0 \\ \hline
\end{tabular}
\caption{Differential inclusive $\ks$ production cross-sections
$\dseta$ for $\pt(\ks)>1$~GeV$/c$ and $|\eta(\ks)|<1.5$ 
in the $W$ range $10<W<125$~GeV. 
The first error is the statistical error and
the second error is the systematic error.}
\label{tab-ks2}
\end{center}
\end{table}

\begin{table}[htbp]
\begin{center}
\begin{tabular}{|c||c|c||c|c|} 
\hline 
 & \multicolumn{2}{c||}{$10<W<35$~GeV} &  \multicolumn{2}{c|}{$35<W<125$~GeV} 
\\ \cline{2-5}
 $\pt$ [GeV/c]     & $\langle \pt \rangle$ [GeV/c] &
d$\sigma$/d$\pt$ [pb/GeV/c] & $\langle \pt \rangle$ [GeV/c] &
d$\sigma$/d$\pt$ [pb/GeV/c] 
\\ \hline 
1.0--1.2 & 1.08 $\pm$0.01 &  127.3  $\pm$  14.6 $\pm$ 12.6    &1.10 $\pm$0.01 & 75.0 $\pm$ \pz9.5 $\pm$ 5.1 \\
1.2--1.5 & 1.33 $\pm$0.01 & \pz73.6 $\pm$ \pz8.4 $\pm$ \pz6.6 &1.33 $\pm$0.01 & 30.0 $\pm$ \pz4.3 $\pm$ 2.3 \\
1.5--1.9 & 1.66 $\pm$0.02 & \pz19.8 $\pm$ \pz3.4 $\pm$ \pz2.5 &1.67 $\pm$0.02 & 13.1 $\pm$ \pz3.0 $\pm$ 2.3\\
1.9--2.4 & 2.11 $\pm$0.03 & \pzz9.8 $\pm$ \pz2.8 $\pm$ \pz1.3 &2.11 $\pm$0.03 &\pz4.4 $\pm$ \pz1.2 $\pm$ 0.4\\
2.4--3.0 & 2.72 $\pm$0.04 & \pzz1.7 $\pm$ \pz0.9 $\pm$ \pz0.2 &2.62 $\pm$0.05 &\pz2.9 $\pm$ \pz0.8 $\pm$ 0.6\\
3.0--4.0 & 3.37 $\pm$0.15 & \pzz0.9 $\pm$ \pz0.4 $\pm$ \pz0.3 &3.38 $\pm$0.09 & \pz1.0 $\pm$ \pz0.6 $\pm$ 0.2\\
4.0--5.5 & \multicolumn{1}{c|}{\hspace{4mm}--} 
& \multicolumn{1}{c||}{\hspace{4mm}--}& 4.56 $\pm$0.15 & \pz0.3 $\pm$ \pz0.3 $\pm$ 0.1\\
\hline
\end{tabular}
\caption{Differential inclusive $\ks$ production cross-sections
$\dspt$ for $\pt(\ks)>1$~GeV$/c$ and $|\eta(\ks)|<1.5$ 
in the $W$ ranges  $10<W<35$~GeV and $35<W<125$~GeV.
The first error is the statistical error and
the second error is the systematic error.} 
\label{tab-ks3}
\end{center}
\end{table}

\clearpage


\end{document}